\documentclass[preprint]{aastex}

\begin{document}

\title{Albedos of Main-Belt Comets 133P/Elst-Pizarro and 176P/LINEAR
        \footnote{
        This work makes use of observations made with the Spitzer Space
        Telescope (Programs 3119 and 30678), which is operated by the Jet Propulsion Laboratory,
        California Institute of Technology under a contract with the National
        Aeronautics and Space Administration (NASA).
        Additionally, some data presented herein were obtained at the W. M.
        Keck Observatory, which is operated as a scientific partnership
        among the California Institute of Technology, the University of
        California, and NASA,
        and was made possible by the generous financial support
        of the W. M. Keck Foundation.  Some data presented herein were also obtained
        at ESO facilities at La Silla under program ID 081.C-0822(A).
        }
}

\author{Henry H. Hsieh$^{a}$, David Jewitt$^{b}$, and
         Yanga R. Fern\'andez$^{c}$}

\affil{
    $^{a}$Astrophysics Research Centre, Queen's University,
    Belfast, BT7 1NN, United Kingdom\newline
    $^{b}$Institute for Astronomy, University of Hawaii, 2680 Woodlawn
    Drive, Honolulu, HI 96822, USA\newline
    $^{c}$Univ. of Central Florida, M.A.P. Building, 4000 Central
    Florida Blvd., Orlando, FL 32816, USA\newline
}
\email{h.hsieh@qub.ac.uk, jewitt@ifa.hawaii.edu, yfernandez@physics.ucf.edu}

\slugcomment{Accepted by ApJL, 2009-02-20}

\begin{abstract}
We present the determination of the geometric $R$-band albedos of
two main-belt comet nuclei based on data from the Spitzer
Space Telescope and  a number of ground-based optical facilities.
For 133P/Elst-Pizarro, we find an albedo of
$p_R=0.05\pm0.02$ and an effective radius of $r_e=1.9\pm0.3$~km
(estimated semi-axes of $a\sim2.3$~km and $b\sim1.6$~km).  For
176P/LINEAR, we find an albedo of $p_R=0.06\pm0.02$ and an
effective radius of $r_e=2.0\pm0.2$~km (estimated semi-axes of
$a\sim2.6$~km and $b\sim1.5$~km).  In terms of albedo, 133P and
176P are similar to each other and are typical of other
Themis family asteroids, C-class asteroids, and other comet nuclei.
We find no indication that 133P and 176P are compositionally
unique among other dynamically-similar (but inactive) members of the 
Themis family, in agreement with previous assertions that the two
objects most likely formed in-situ.
We also note that low albedo ($p_R<0.075$) remains a consistent
feature of all cometary ({\it i.e.}, icy) bodies, whether they originate in the
inner solar system (the main-belt comets)
or in the outer solar system (all other comets).
\end{abstract}


\newpage

\section{INTRODUCTION}

The main-belt comets (MBCs), of which 133P/Elst-Pizarro (hereafter, 133P)
and 176P/LINEAR (hereafter, 176P) are examples,
occupy stable orbits that are decoupled from Jupiter
and which are indistinguishable from the orbits of other main-belt
asteroids \citep{hsi06b}.  Dynamical simulations show that MBCs
are extremely unlikely to originate in the Kuiper Belt given the current
configuration of the major planets \citep[{\it e.g.},][]{jfer02},
indicating that
they are instead likely to be native to the main asteroid belt.
Recent work suggests that some icy Kuiper Belt objects
might have been delivered to the asteroid belt
during the Late Heavy Bombardment \citep{lev08}, but even
those simulations fail to produce the low-inclination,
low-eccentricity orbits of MBCs such as 133P and 176P.

In this letter, we use observations from the Spitzer Space Telescope
\citep[hereafter, Spitzer;][]{wer04} to determine the geometric albedos of 133P and
176P, and then discuss the implications of these measurements.

\section{OBSERVATIONS}

We obtained optical observations of 133P and 176P on multiple
occasions from 2003 through 2008 using the 10~m Keck I and University of
Hawaii (UH) 2.2~m telescopes on Mauna Kea, and the 3.58~m New Technology
Telescope (NTT) at the European Southern Observatory (ESO) at La Silla.
Observations with the UH 2.2~m telescope were made using either a
Tektronix 2048$\times$2048 pixel CCD or the Orthogonal Parallel Transfer
Imaging Camera \citep[OPTIC;][]{ton04},
both behind standard Kron-Cousins BVRI broadband filters.
Observations with Keck were made using the Low Resolution Imaging
Spectrometer \citep[LRIS;][]{oke95} in imaging mode.  LRIS employs a
Tektronix $2048\times2048$ CCD with standard Kron-Cousins BVRI filters.
Observations with the NTT were made using the ESO Faint Object
Spectrograph and Camera \citep[EFOSC2;][]{buz84}, which
employs a 2048$\times$2048~pixel Loral/Lesser CCD behind Bessel BVR
broadband filters.

Bias subtraction and flat-field reduction
were performed for all optical data.  
Dithered images of the twilight sky were used to construct flat fields
for UH 2.2~m data, while
images of the illuminated interior of the telescope dome were used to
construct flat fields for Keck and NTT data.
Photometry of our target objects and \citet{lan92} standard
stars was obtained by measuring net fluxes within circular
apertures of varying radii depending on the nightly seeing,
with background sampled from surrounding circular annuli.

\begin{figure}
\plotone{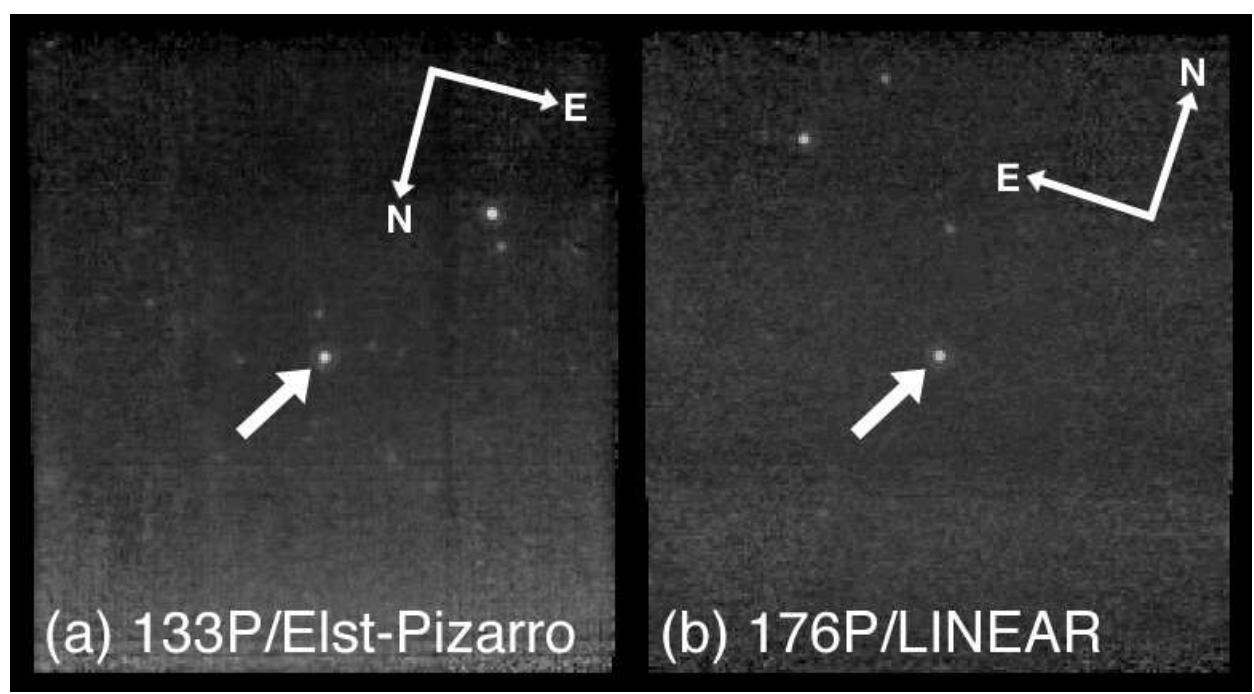} 
\caption{\small Composite (PBCD) 24~$\mu$m images of (a) 133P/Elst-Pizarro (166~s
                total exposure time) and (b) 176P/LINEAR (48~s total exposure time),
		indicated by arrows, obtained using MIPS on Spitzer.  Both objects
		are point sources with no indication of cometary activity.
		Each panel is $\sim7\farcm5$ by $8\farcm2$ in size.
}
\label{spitzer_images}
\end{figure}

Spitzer observations of 133P (3 visits,
166~s of total exposure time per visit; Fig.~\ref{spitzer_images}a),
using the 24~$\mu$m channel (effective wavelength of 23.68~$\mu$m)
of the Multiband Imaging Photometer for Spitzer \citep[MIPS;][]{rie04}
and originally obtained on 2005 April 11 as part of Cycle 1 program
3119 \citep{rea07}, were retrieved from the Spitzer archive.
Observations of 176P
(2 visits, 48~s of total exposure time per
visit; Fig.~\ref{spitzer_images}b), also with the 24~$\mu$m channel of MIPS, were obtained on
2007 January 1 as part of Cycle 3 program 30678.  Observational
circumstances are shown in Table~\ref{obs_spitzer}.
Photometry of our target objects from pipeline-processed
Spitzer post-Basic Calibrated Data (PBCD) was obtained by measuring net
fluxes within circular apertures with 6-pixel ($14\farcs7$) radii,
and then applying appropriate aperture corrections (1.14 in the case of
a 6-pixel aperture) and color corrections (0.96 for both targets).

\section{RESULTS\label{results}}

\begin{figure}
\plotone{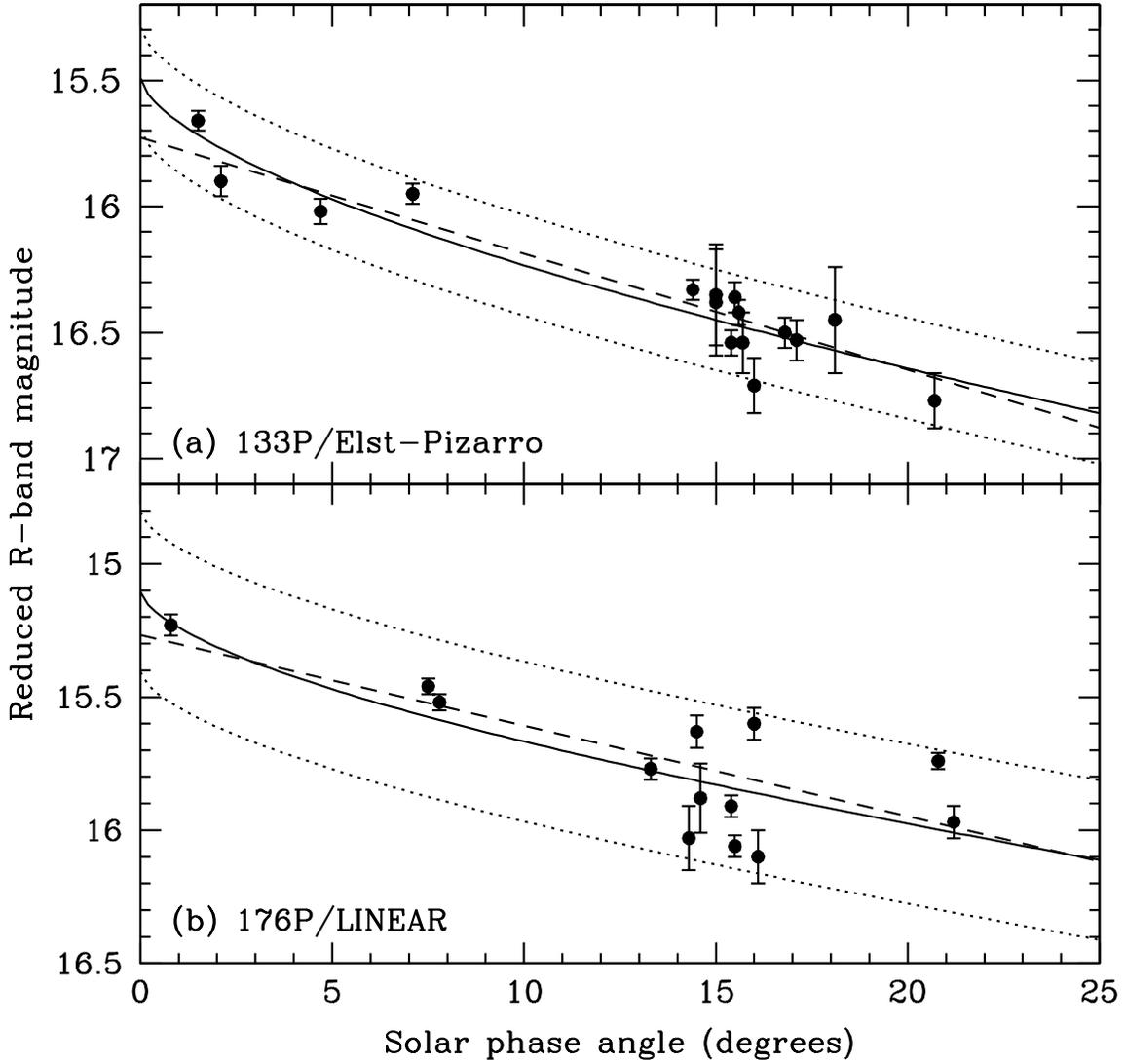} 
\caption{\small Phase function solutions for (a) 133P/Elst-Pizarro and
                 (b) 176P/LINEAR, with best-fit IAU phase laws
                 plotted as solid lines and best-fit linear phase
		 functions plotted as dashed lines.  Dotted lines
		 indicate the expected range of brightness deviations
		 from the IAU phase law due to rotation of the body.
		 Observed reduced $R$-band magnitudes are plotted as
		 solid circles.
}
\label{phaselaws}
\end{figure}

We use our optical data to find best-fit IAU phase function parameters
for 133P of $H_R=15.49\pm0.05$~mag and $G=0.04\pm0.05$, and best-fit
linear phase function parameters (omitting data obtained
at solar phase angles at which opposition surge effects are expected) of
$m_R(1,1,0)=15.69\pm0.05$~mag and $\beta=0.049\pm0.004$~mag~deg$^{-1}$.
These parameters are calculated using photometry obtained while
133P was observed to be inactive, and as such, are a refinement
of parameters previously derived by \citet{hsi04}
from photometry obtained while 133P was visibly active.
For 176P, we find corresponding parameters of
$H_R=15.10\pm0.05$~mag, $G=0.26\pm0.05$, $m_R(1,1,0)=15.27\pm0.05$~mag,
and $\beta=0.034\pm0.005$~mag~deg$^{-1}$.  These parameters were likewise
calculated only using photometry obtained while the comet was observed to
be inactive.  Plots of phase function
solutions for both objects are shown in Figure~\ref{phaselaws}.
From their phase functions, we estimate our targets' expected mean optical
magnitudes as viewed from Spitzer at the time of their observations
to be $m_R=21.63$~mag for 133P and $m_R=20.39$~mag for 176P.

Both objects exhibit significant rotational brightness variations,
however, which represent significant sources of uncertainty in the
interpretation of our infrared data.
A rotation period of $P_{rot}=3.471$~hr and a lightcurve range of
$\Delta m=0.4$~mag have been previously found for 133P \citep{hsi04}.
The rotational properties of 176P are currently poorly constrained.
On 2007 March 21, however, we observed a photometric range for the
object of $\Delta m \approx 0.6$~mag over $\sim4.5$~hr,
suggesting a rotation period of $P_{rot}\geq18$~hr
(assuming a double-peaked lightcurve).
This is consistent with \citet{lic07a} who found $P_{rot}>22$~hr.

Fortunately, constraints on the rotational phase of each object
can be derived from the infrared data.
Our second flux density measurement for 176P was 1.4 times larger
than the first, implying an equivalent increase in visible
cross-sectional area, corresponding to a change in visual
magnitude of $\Delta m = -0.37$~mag, a significant fraction of
the object's inferred optical photometric range.
Such a large magnitude change indicates
that the object was necessarily first observed near the minimum
and then near the maximum of its lightcurve (consistent with
the 4.55~hr interval between the two observations).  Using
these constraints, we are able to adjust our optical brightness
estimates (Table~\ref{obs_spitzer}) accordingly, thereby
reducing the effects of rotational phase uncertainty.
The three Spitzer observations of 133P span only 7~min meaning that
the rotational phase is less tightly constrained.
Given 133P's short rotation period, however, a small decrease in the
scattering cross-section is still detectable between the first and
last Spitzer observation, corresponding to a change in visual
magnitude of $\Delta m = 0.14$~mag and leading to the revised optical
brightness estimates for 133P in Table~\ref{obs_spitzer}.

We use the \citet{har98} Near-Earth Asteroid Thermal Model (NEATM) to
iteratively solve for the effective radius, $r_e$, and geometric $R$-band
albedo, $p_R$, of each object.  As with any model, NEATM requires a
number of assumptions, which, in turn, introduce uncertainties.  One such
source of uncertainty is the
phase effect for thermal emission.  NEATM treats the effect
geometrically, calculating it based on the fraction of the Earth-facing
hemisphere that is illuminated by the Sun at the time of observation.
While this effect has been poorly measured and thus poorly constrained
for large phase angles, the infrared phase coefficient of
0.01~mag~deg$^{-1}$ that we use here is generally
considered to be appropriate for phase angles
$\alpha<30^{\circ}$ \citep[{\it cf}.][]{mor77,har98}.  Thus, given
the small phase angles ($14^{\circ}<\alpha<17^{\circ}$)
at which the Spitzer observations were obtained,
this effect should introduce minimal
systematic uncertainty into our calculations.

A more significant issue is that of the
beaming parameter, $\eta$.
We lack the minimum number of data points needed to constrain $\eta$
for either 133P or 176P, forcing us to assume its value.
A Spitzer survey of $\sim50$ Jupiter-family comet nuclei by
\citet{fer08}, however, found values of $0.6<\eta<1.2$,
and all were consistent with $\eta\approx0.94\pm0.20$.
Given the results of this survey
and assuming that 133P and 176P have low thermal inertias
\citep[similar to other comet nuclei, {\it e.g.}, 9P/Tempel 1,
which has $I<50$~W~K$^{-1}$~m$^{-2}$~s$^{1/2}$;][]{gro07},
we adopt $\eta=1.0$
as a reasonable assumption for solving for $r_e$ and $p_R$.
To account for uncertainties in $\eta$, we also perform
parallel calculations for $\eta=0.8$ and $\eta=1.2$ (Table~\ref{albedos}).

Thus, assuming an emissivity of $\varepsilon=0.9$,
we find $r_e=1.9\pm0.3$~km and $p_R=0.05\pm0.02$ for
133P, and $r_e=2.0\pm0.2$~km and $p_R=0.06\pm0.02$ for 176P.
Estimated errors for both objects are mainly due to
uncertainties in both $\eta$ and rotational phase.
Given the observed photometric ranges ($\Delta m_{\rm 133P}\approx0.40$~mag;
$\Delta m_{\rm 176P}\approx0.60$~mag) and corresponding inferred minimum axis
ratios ($[a/b]_{\rm 133P}\approx1.45$; $[a/b]_{\rm 176P}\approx1.74$) for each object,
we find $a\sim2.3$~km and $b\sim1.6$~km for 133P, and
$a\sim2.6$~km and $b\sim1.5$~km for 176P
as our best estimates of the semiaxes of each object.

\section{DISCUSSION\label{discussion}}

\begin{figure}
\plotone{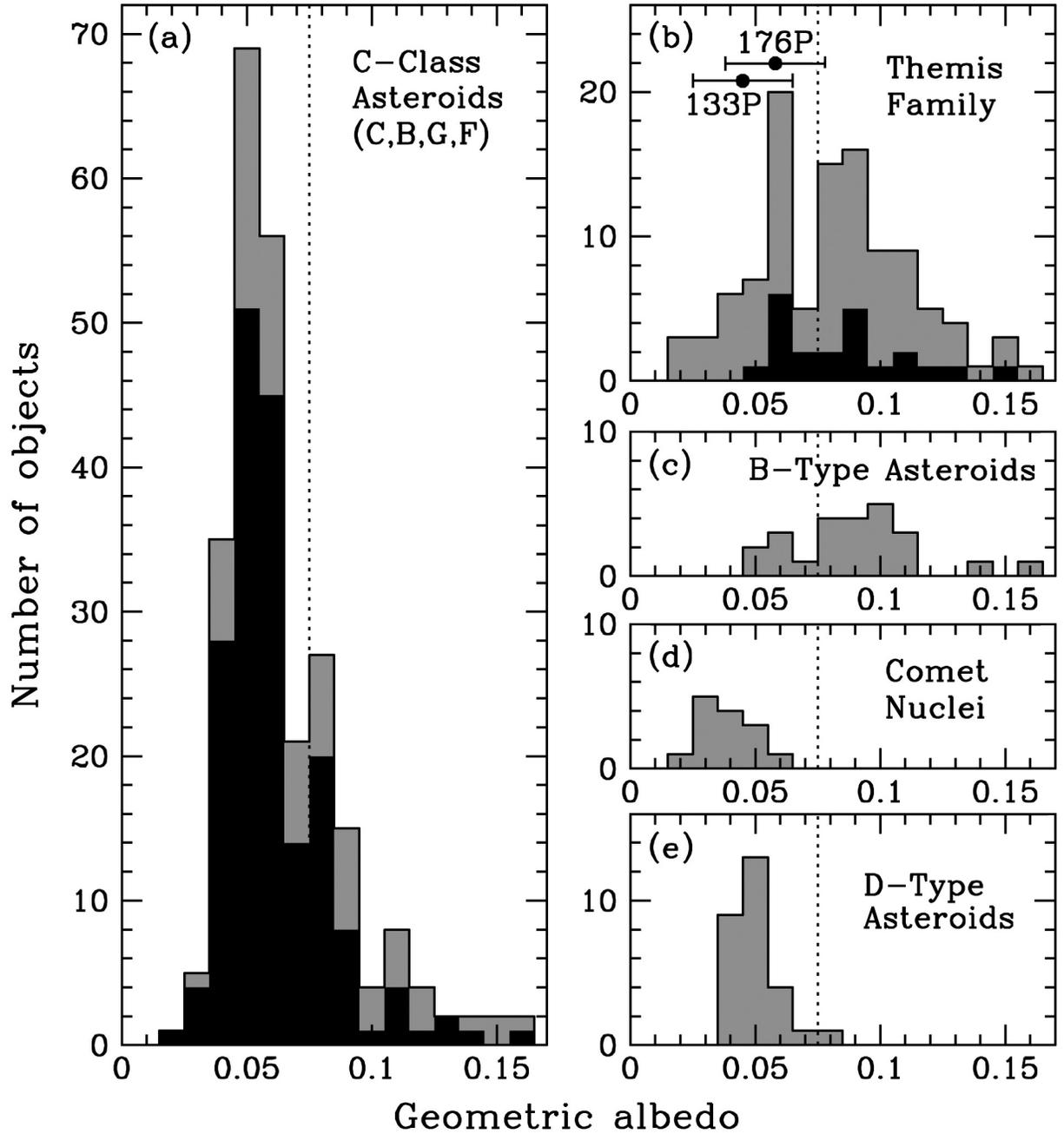} 
\caption{\small Histograms showing albedo distributions for (a) C-class
                 (C-, B-, G-, or F-type) asteroids, where the superimposed
		 black-shaded histogram only includes objects explicitly
		 classified as C-type asteroids,
		 (b) dynamical members of the Themis asteroid family, where
		 the superimposed black-shaded histogram only includes those
		 Themis members with measured albedos that have been
		 classified as C-class asteroids (not all
		 have been assigned taxonomic classes to date, however),
		 (c) B-type asteroids,
		 (d) active comet nuclei, and
		 (e) D-type asteroids.
		 All taxonomic classifications follow the Tholen
		 system \citep{tho89,laz04}.  Comet
		 nucleus albedos are from \citet{lam04} and \citet{bro04},
		 while all other albedo values are from the IRAS Minor Planet
		 Survey \citep{ted04}.  Objects with $p<0.075$, designated as
		 ``cometary'' by \citet{fer05}, are to the left of
		 the dotted lines.
}
\label{albedohists}
\end{figure}

We plot histograms showing the albedo ($p_V$) distributions of several
solar system body populations of interest in Figure~\ref{albedohists}.
Assuming that both 133P and 176P are approximately spectrally neutral
({\it i.e.}, $p_V\approx p_R$), based on 133P's spectral classification
as a C- or B-type asteroid and 176P's classification as a B-type
asteroid \citep{lic07a}, we find that their albedos are typical of
C-class asteroids (Fig.~\ref{albedohists}a) and are also well within
the distribution of albedos measured for members of the Themis asteroid
family (Fig.~\ref{albedohists}b), with which 133P and 176P appear to be
dynamically associated \citep{hsi06b}.

The Themis family is dominated by C-class asteroids
\citep[{\it cf}.][]{flo99}, of which a substantial fraction
(17 of the 39 currently classified members of the family, or $\sim44$\%)
belong to the subclass of B-type asteroids.
For comparison, B-type asteroids comprise only $\sim15$\%
of the general C-class population and $\sim5$\% of all currently
classified asteroids \citep{tho89,laz04}.  In terms of albedo, we find
133P and 176P to be consistent with both C-type asteroids and B-type
asteroids ({\it cf.} Figs.~\ref{albedohists}a and \ref{albedohists}d),
in agreement with their spectral classifications by \citet{lic07a}.  
Thus, in terms of both albedo and spectral type,
133P and 176P appear to be typical Themis asteroids, supporting
previous speculation that the family might be home to more MBCs \citep{hsi06a}.

Other objects like 133P and 176P that also have orbits considered to be
dynamically asteroidal, yet have been associated with observed or inferred
cometary activity, are also classified as C-class objects. One such
object is the cross-listed comet-asteroid 107P/(4015) Wilson-Harrington
\citep[C- or F-type;][]{tho89}.  Other examples include Geminid meteor
stream parent 3200 Phaethon \citep[B- or F-type;][]{tho89,lic07b}
and its possible fragment, 155140 (2005 UD) \citep[B- or F-type;][]{jew06,kin07}.
Another likely fragment of 3200 Phaethon, 1999 YC, appears spectrally neutral
and is classified as a C-type object \citep{kas08}.

Being below the upper bound of ``comet-like'' albedos ($p_R=0.075$)
employed by \citet{fer05}, the albedos of 133P and 176P are also
consistent with those of the nuclei of other active comets
(Fig.~\ref{albedohists}d).  Spectroscopically, comet nuclei exhibit
a broad range of colors, with both D-type-like and C-type-like
spectral reflectivity gradients being found for various comets
\citep[{\it cf}.][]{fit94,jew02}, and in terms of albedos, the two
MBC nuclei we consider here are consistent with both spectral
types (Fig.~\ref{albedohists}a \& \ref{albedohists}e).  Thus, despite
their strong dynamical association with main-belt asteroids, we find that
133P and 176P have surfaces that may be compositionally comparable
to other comets.  This is consistent with
\citet{jew02} who suggested that the surface properties of short-period
comet nuclei were likely largely due to sublimation-driven evolutionary
effects and were not primordial in nature.

Studying the surface properties of the main-belt comets is vital 
for understanding their evolution and putting their volatile content into
the proper context.
In light of those goals, we find that, in terms of albedos,
(1) 133P and 176P are similar to each other,
(2) they are typical of other Themis asteroids and the C- and B-type
    asteroids that dominate the Themis family, and
(3) their albedos are also consistent with albedos measured for other
    comet nuclei and D-type asteroids.
Given these results, we find that low albedo continues to be a consistent
feature of all cometary bodies, whether they originate in the outer or
inner solar system.  This finding necessarily also means,
however, that albedo does not appear to be an effective diagnostic of
the region from which a comet originates.

\begin{acknowledgements}
We acknowledge support of this work through STFC fellowship grant ST/F011016/1 to HHH,
NASA Spitzer grant JPL-1289078 and NASA Planetary Astronomy grant NNG05GF76G to DJ,
and NASA grant JPL-1289123 to YRF.  We also thank Bill Reach for valuable discussion
and Alan Harris (DLR, Berlin) for pointing out an error in our initial albedo calculations.
\end{acknowledgements}

\begin{deluxetable}{llccrrc}
\tablewidth{0pt}
\tablecaption{Spitzer Observations\label{obs_spitzer}}
\tablecolumns{7}
\tablehead{
\colhead{Object} & \colhead{Date}
   & \colhead{UT}
   & \colhead{$R$\tablenotemark{a}}
   & \colhead{$\Delta_{Sp}$\tablenotemark{b}}
   & \colhead{$\alpha_{Sp}$\tablenotemark{c}}
   & \colhead{$m_R$\tablenotemark{d}}
}
\startdata
133P/Elst-Pizarro & 2005 Apr 11 & 08:01:11 & 3.596 & 3.046 & 14.6 & $21.56\pm0.15$ \\ 
                  & 2005 Apr 11 & 08:04:49 & 3.596 & 3.046 & 14.6 & $21.63\pm0.15$ \\ 
                  & 2005 Apr 11 & 08:08:30 & 3.596 & 3.046 & 14.6 & $21.70\pm0.15$ \\ 
176P/LINEAR       & 2007 Jan 01 & 00:49:12 & 3.162 & 2.541 & 16.2 & $20.58\pm0.11$ \\ 
                  & 2007 Jan 01 & 05:22:13 & 3.163 & 2.539 & 16.1 & $20.21\pm0.11$ \\ 
\enddata
\tablenotetext{a} {Heliocentric distance in AU}
\tablenotetext{b} {Distance from Spitzer in AU}
\tablenotetext{c} {Solar phase angle (Sun-object-Spitzer) in degrees}
\tablenotetext{d} {Expected $R$-band magnitude as calculated from
                    rotational-phase information inferred from the infrared data and
                    observationally-determined $H,G$ phase functions (Fig.~\ref{phaselaws}).
		    Listed errors account for uncertainties in both rotational
		    phase and phase function solutions.
                  }
\end{deluxetable}

\begin{deluxetable}{lrccrccrcc}
\tablewidth{0pt}
\tablecaption{Albedos and Radii Computed from Optical and Infrared Observations\label{albedos}}
\tablecolumns{10}
\tablehead{
 & \multicolumn{3}{c}{$\eta=0.8$\tablenotemark{a}}
 & \multicolumn{3}{c}{$\eta=1.0$\tablenotemark{a}}
 & \multicolumn{3}{c}{$\eta=1.2$\tablenotemark{a}} \\
\colhead{Object}
  & \colhead{$F_{24\mu m}$} & \colhead{$r_e$} & \colhead{$p_R$}
  & \colhead{$F_{24\mu m}$} & \colhead{$r_e$} & \colhead{$p_R$}
  & \colhead{$F_{24\mu m}$} & \colhead{$r_e$} & \colhead{$p_R$}
}
\startdata
 133P/Elst-Pizarro &  $6.4\pm0.1$ & 1.78 & 0.054 &  $6.5\pm0.1$ & 1.94 & 0.045 &  $6.5\pm0.1$ & 2.09 & 0.039 \\
                   &  $6.0\pm0.1$ & 1.72 & 0.054 &  $6.0\pm0.1$ & 1.88 & 0.045 &  $6.1\pm0.1$ & 2.03 & 0.039 \\
                   &  $5.7\pm0.1$ & 1.67 & 0.054 &  $5.7\pm0.1$ & 1.83 & 0.045 &  $5.7\pm0.1$ & 1.97 & 0.039 \\
 176P/LINEAR       & $10.4\pm0.2$ & 1.72 & 0.068 & $10.5\pm0.2$ & 1.87 & 0.058 & $10.5\pm0.2$ & 2.01 & 0.050 \\
                   & $14.6\pm0.2$ & 2.04 & 0.068 & $14.7\pm0.2$ & 2.22 & 0.058 & $14.7\pm0.2$ & 2.38 & 0.050 \\
\enddata
\tablenotetext{a} {Assumed $\eta$ value used to compute aperture- and color-corrected 24~$\mu$m
                    flux ($F_{24\mu m}$) in mJy (where uncertainties are estimated from sky background statistics),
		    effective radius ($r_e$) in km, and geometric
		    $R$-band albedo ($p_R$)}
\end{deluxetable}

\clearpage

\clearpage

\clearpage

\end{document}